\documentclass[twocolumn]{aastex701}

\usepackage{caption,subcaption}
\usepackage[english]{babel}
\usepackage{graphicx}
\usepackage{amsmath}
\usepackage{natbib}
\usepackage{gensymb}
\usepackage{multirow}
\usepackage{xcolor}
\usepackage{soul}
\usepackage{cancel}

\hypersetup{linkcolor=red,citecolor=green,filecolor=cyan,urlcolor=magenta}

\begin{document}

\title{Flux Dependence of the 1\,keV Soft X-ray Feature in SAX J1808.4-3658 with \textit{NICER}}

\author[]{Katherine Bruce}
\affiliation{Department of Physics, Montana State University, 
Bozeman, MT 59717, USA}\email[show]{katherinebruce@montana.edu}  

\author[]{Anne Lohfink}
\affiliation{Department of Physics, Montana State University, 
Bozeman, MT 59717, USA}\email{anne.lohfink@montana.edu}

\author[]{Jaiverdhan Chauhan}
\affiliation{School of Physics and Astronomy, University of Leicester, University Road, Leicester LE1 7RH, UK}\email{jc999@leicester.ac.uk}

\author[]{Suchismito Chattopadhyay}
\affiliation{Inter-University Centre for Astronomy and Astrophysics, Ganeshkhind, Pune 411007, India}\email{suchismito.chattopadhyay@iucaa.in}

\author[]{Sachiko Tsuruta}
\affiliation{Department of Physics, Montana State University, 
Bozeman, MT 59717, USA}\email{tsuruta@montana.edu}
\affiliation{Kavli Institute for the Physics and Mathematics 
of the Universe (WPI), The University of Tokyo, Kashiwa, Chiba 
277-8583, Japan}

\author[]{Andrew C. Liebmann}
\affiliation{Department of Physics, Montana State University, 
Bozeman, MT 59717, USA}\email{bloodcorba@earthlink.net}

\begin{abstract}
We present a systematic X-ray spectral analysis of the 2019 and 2022 outbursts of the accreting millisecond pulsar SAX J1808.4-3658 with \textit{NICER} with a particular emphasis on the nature of the longstanding $\sim$1\,keV spectral feature. We report here the first clear evidence that the 0.9–1.0 keV feature varies systematically with source brightness, exhibiting strong correlations with both continuum and blackbody flux. This establishes that the feature is directly coupled to the accretion process. The scaling and energy distribution of the feature differ between outbursts, corresponding to a $\sim$2\,dex change in reprocessor density (log\,$n_e \approx 19.1$ in 2019 vs. $\sim$17.3 in 2022), which naturally explains the stronger, broader feature in 2019 and the weaker, more localized $\sim$1\,keV feature in 2022, in agreement with high-density reflection models. These results provide the strongest constraints to date on the origin of the $\sim$1\,keV feature in SAX J1808.4-3658.

\end{abstract}

\section{Introduction}\label{sec:intro}

A neutron star can be spun up through sustained mass transfer from a low-mass companion, evolving into an accreting millisecond X-ray pulsar (AMXP; see \cite{disalvo2022}, \cite{patruno2021}, \cite{campana2018} for reviews), where the interplay between accretion and strong-field relativistic effects offers insight into accretion processes, relativity, and high-density physics. SAX J1808.4–3658 (hereafter J1808) was the first AMXP discovered and remains the best studied example, due to its frequent outbursts (1996, 1998, 2002, 2005, 2008, 2011, 2015, 2019, 2022, and 2025) \citep{zand,wijnands,markwardt02,greenhill,markwardt08,patruno2012, bult, illiano23, atel25}.

Many important parameters have been constrained for this system, including a spin frequency of 401\,Hz and orbital period of 2.01 hr \citep{chakrabarty}. During many outbursts, type-I thermonuclear X-ray bursts have been detected in the system \citep{galloway, galloway08, bult}. The neutron star's companion is thought to be a brown dwarf with a mass of approximately 0.05 M$_\odot$ \citep{bildsten}. Although classified as an atoll source, J1808 exhibits atypical behavior, primarily inhabiting the island (hard) state and only rarely transitioning to the extreme island (hardest) or lower left banana (soft) states \citep{straaten,bult2015}. Distance estimates place the system at  2.7$\pm$ 0.3\,kpc, and modeling of thermonuclear bursts suggests a neutron star mass of 0.9-2.1 M$_\odot$ and a radius of 10.0-13.1\,km \citep{galloway2024,goodwin2019}. 

A long‑standing question in accreting neutron‑star systems is the origin of a soft X‑ray excess near 1\,keV. This feature is common among neutron‑star low mass X‑ray binaries (LMXBs) and has been observed repeatedly in J1808 using several different X-ray missions (e.g. \cite{bult,sharma23,casten23, kaushik25}). High‑resolution spectroscopy has associated similar features with Fe L‑shell and O K‑shell emission from a photoionized disk corona \citep{schulz,asai,cackett2010}, while disk reflection may also contribute \citep{fabian2000,chakraborty}. 

A shift in the centroid energy of the 1\,keV feature between the 2019 and 2022 outbursts in J1808 was first identified in a study that analyzed only the joint \textit{NICER} and \textit{NuSTAR} observations \citep{bruce}. This initial result, based on a limited number of observations, raised the possibility that the soft‑band feature evolves systematically over time and motivated the deeper investigation carried out here. 

During the 2019 and 2022 outbursts \textit{NICER} monitored J1808, providing an ideal data set for testing how the 1\,keV feature depends on flux and other spectral parameters. In this paper, we use these observations to track the evolution of the soft X‑ray spectrum throughout both outbursts and into the reflaring phase. These data allow us to examine whether the strength and centroid energy of the 1\,keV excess vary systematically with source brightness, accretion state, or changes in other spectral parameters. As we show below, the feature becomes stronger at higher fluxes.

In Section \ref{sec:red}, we describe the \textit{NICER} data preparation and generation of X-ray spectra. Section \ref{sec:results} details our spectral modeling, the presence of the 1\,keV feature, and its dependence on other source properties. We discuss the implications of these results in Section~\ref{sec:disc}.
 
\section{Observations and Data Reduction}\label{sec:red}

From 2019 July 30 to 2019 November 8, NICER obtained 63 pointed observations of the source, followed by an additional 53 observations collected between 2022 August 19 and 2022 October 31. Some of these data have been previously analyzed, first by \cite{bult} for the 2019 dataset and by \cite{illiano23} for the 2022 dataset. All data were reduced using HEASoft v6.35.1, including NICERDAS v13 and XSELECT v3.0, XSPEC v12.15.0, and pyXspec v2.1.4 \citep{blackburn,arnaud}. Level‑3 spectral products were generated with the nicerl3‑spect pipeline using initially standard \textit{NICER} screening criteria, and adopting the SCORPEON v23 background model\footnote{\url{https://heasarc.gsfc.nasa.gov/docs/nicer/analysis_threads/scorpeon-xspec/}}. For each observation, the standard SCORPEON background components were allowed to vary, along with the source wind components for the O K, O VII, and O VIII lines. The light curves were extracted using nicerl3‑lc with a temporal resolution of 1.0 s. Several non-X-ray flares were identified in both datasets, so times when the cut-off rigidity was in the range of 0-1.5 GeV/c and the overshoot range exceeded 5\,ct/s were excluded following the advice given on the \textit{NICER} website\footnote{\url{https://heasarc.gsfc.nasa.gov/docs/nicer/analysis_threads/flares/}}. The two type-I thermonuclear bursts identified in \cite{bult} and \cite{casten23} were also removed for analysis. To ensure that the source appears above the background, any observation with an exposure of less than 250 s or an average rate less than 15\,ct/s was discarded. 

Spectral analysis was restricted to the 0.7–10.0\,keV energy range. Below approximately 0.7\,keV, the data showed a repeatable, non‑astrophysical feature spanning 0.22–0.65 keV. This low‑energy structure did not correlate with source behavior and is likely associated with time‑variable detector response or residual background systematics. To avoid introducing spurious spectral curvature, these channels were excluded from all spectral modeling. Spectra were grouped to a minimum of 20 counts per bin to ensure the validity of $\chi^{2}$ statistics in XSPEC. The data preparation yielded 81 high-quality spectral results with an average exposure time of 4.8\,ks to study the soft X-ray spectral evolution of the 2019 and 2022 outbursts and the properties of the 1\,keV feature.  

\section{Results}\label{sec:results}
\subsection{Continuum Modeling}
\begin{figure*}
    \centering
    \includegraphics[width=\linewidth]{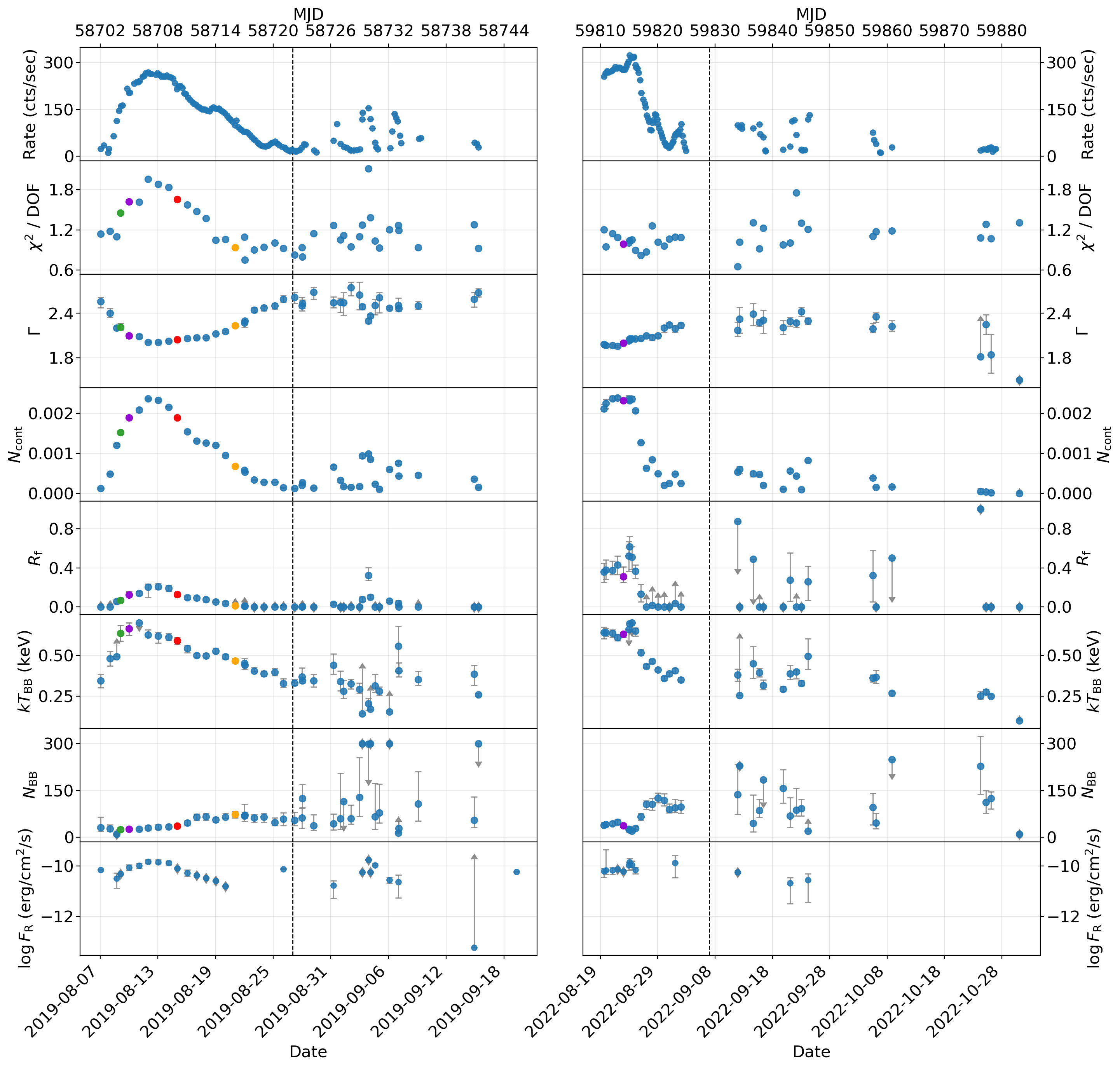}
        \caption{Evolution of SAX~J1808.4$-$3658 across the 2019 and 2022 outbursts as observed by \textit{NICER}. 
Panels show (from top to bottom) the 0.7--10~keV count rate (per observation), the best fit $\chi^{2}$,the best-fit continuum values for photon index ($\Gamma$), continuum normalization (N$_{cont}$), reflection fraction (R$_{f}$), blackbody temperature (kT$_{BB}$; keV), and blackbody normalization (N$_{BB}$) from the baseline model fitted \emph{without} a Gaussian line at 1~keV (\texttt{TBfeo} $\times$ [\texttt{relxillCp} + \texttt{bbodyrad}]; see Section~\ref{sec:results}), and logarithmic reflection flux (log F$_{R}$; erg/cm$^2$/s). 
Only the photon index, reflection fraction, blackbody temperature, and component normalizations were allowed to vary during spectral modeling; all other parameters were fixed to values adopted from \citet{bruce} (See Table \ref{tab:params} in Appendix A). 
The vertical dashed line marks the onset of the reflaring phase \citep{baglio2020,illiano23}. 
Colored markers indicate contemporaneous reference points from prior studies:  green—\citet{casten23}; purple—\citet{bruce}; red-\citet{sharma23}; orange—\citet{bult}. 
Spectra were grouped to a minimum of 20 counts per bin and fitted with $\chi^{2}$ statistics (Section~\ref{sec:red})}
    \label{fig:compplot}
\end{figure*}

In \cite{bruce}, \textit{NuSTAR} observations taken in 2019 and 2022 were modeled alongside the nearest \textit{NICER} observations. This broadband X-ray spectral modeling revealed that the X-ray spectra of J1808 during outburst are well described by a model consisting of photoelectric absorption (\texttt{TBFEO}), a thermally Comptonized primary continuum and reflection component (\texttt{relxillCp}), a blackbody (\texttt{BBODYRAD}), and a Gaussian component near 1\,keV \citep{wilms,zdziarski,garcia}. The centroid of this Gaussian feature was found to differ slightly between outbursts, motivating a closer examination of how the feature behaves within individual outbursts using the more extensive \textit{NICER} coverage.

Our goal in this work is to investigate this 1 \,keV feature further and systematically study its variability. To do this, we first established a continuum baseline by fitting all \textit{NICER} observations without a Gaussian line component, otherwise adopting the same spectral components identified in \cite{bruce}. Not all parameters can be robustly constrained from \textit{NICER} alone, and therefore only a selection of parameters was left free to vary. These variables are the power law index ($\Gamma$), reflection fraction, blackbody temperature (kT$_{\text{bb}}$), and component normalizations. All other parameters were set to values from \cite{bruce} (See Table \ref{tab:params} in Appendix A). While some of these parameters, such as electron temperature or ionization, are expected to change over the course of the outburst, we are not able to constrain them in our modeling due to either the limited spectral energy range or modeling degeneracies. In particular, we tested allowing the ionization parameter and disk density to vary, finding no impact on the results reported in this work.

For each observation, the baseline model yielded an acceptable fit that describes the overall shape of the spectrum well, with a reduced $\chi^2$ of less than 2 for each initial baseline continuum model. The number of degrees of freedom varied between 93 and 150 across observations. Higher reduced $\chi^2$ values, particularly in the brighter portions of the outburst, are typically associated with the prominent 1\,keV feature in the residuals, as will be discussed in more detail later in this section. An overview of the results from this baseline model are shown in Figure \ref{fig:compplot}. 

In 2019, \textit{NICER} captured several observations before the outburst peak, the peak and then the continued slow decline. As previously noted by \cite{baglio2020}, the source remains in the hard state throughout. As the source brightened toward maximum, the photon index hardened (reaching its minimum near the peak) while both the blackbody temperature and reflection fraction increased, then all trended back as the flux declined. The source reaches the end of its decay phase about 20\,days after the observations begin on 2019-08-27 \citep{baglio2020}. Then, the reflaring portion begins. During this period of the outburst, these mini-outburst events cause rapid changes in the system. Two distinct peaks are observed in the light curve during the reflaring. The sudden increase in reflection fraction and blackbody normalization is likely due to one of these reflaring events. Around 2019-09-04, a reflaring event is observed along with a jump in the reduced chi-squared statistic. 

In 2022, \textit{NICER} monitoring began near the outburst peak, followed by a faster return to quiescence than in 2019 and the onset of reflaring roughly two weeks after maximum \citep{illiano23}.
Unlike the 2019 outburst where the continuum hardened towards the peak, the photon index remained relatively constant around the 2022 peak and then softened as the flux dropped. Because 2022 monitoring began near the outburst maximum, the rise phase is not covered, limiting direct comparison. The blackbody temperature peaked contemporaneously with the X‑ray maximum and decreased thereafter, while the blackbody normalization increased during the sharp flux decline (consistent with a larger emitting area in the blackbody component). 

\subsection{Soft Feature}
With this baseline continuum behavior characterized and an acceptable broad-band component model established, each spectrum was then systematically searched for the 1\,keV feature.

A Gaussian component with zero width was added to the baseline model. Around 1\,keV, \textit{NICER}'s spectral resolution is 85\,eV, such that a moderately blended line complex is not spectrally resolved, and the Gaussian serves as a representation of unresolved emission excess. We also tested allowing the Gaussian width to vary, but the width was not meaningfully constrained given \textit{NICER}'s spectral resolution around 1\,keV. All parameters except for the line normalization were held fixed. 
Line energies between 0.7 and 1.4\,keV were tested in steps of 0.1\,keV, chosen to sample the spectrum at a resolution comparable to the instrument response, fitting the line normalization and recording the resulting $\chi^2$ improvement at each step. For each trial energy, a line was considered to be detected if the addition of the Gaussian component improved the result by $\Delta \chi^2\geq 9$, corresponding to a 3$\sigma$ single-trial significance. This threshold represents single-trial significance and does not account for the energies searched; we therefore use it as a detection criterion rather than a global significance estimate. 

Across both outbursts, detections are mainly observed in the 0.9\,keV and 1.0\,keV bins. In 2019, the line feature is absent in the first several observations of the outburst during the rise. Around the peak, it then appears in both the 0.9 and 1.0 keV bins, eventually becoming stronger in the 0.9\,keV bin before declining below the detection limit during the decline. The 0.9\,keV line is then detected fairly strongly around 2019-09-03, after reflaring begins, followed by two much weaker detections closely after.

During the 2022 outburst, detections of this feature are generally weaker. The detection strength peaks near the outburst maximum before fading below the detection limit as rapidly as the luminosity. One detection is observed during the 2022 reflaring period. In summary, the timing of line detections appears systematic rather than random, with the feature recurring alongside the evolution of the outburst. 

\begin{figure*}[th]
    \centering
    \includegraphics[width=\linewidth]{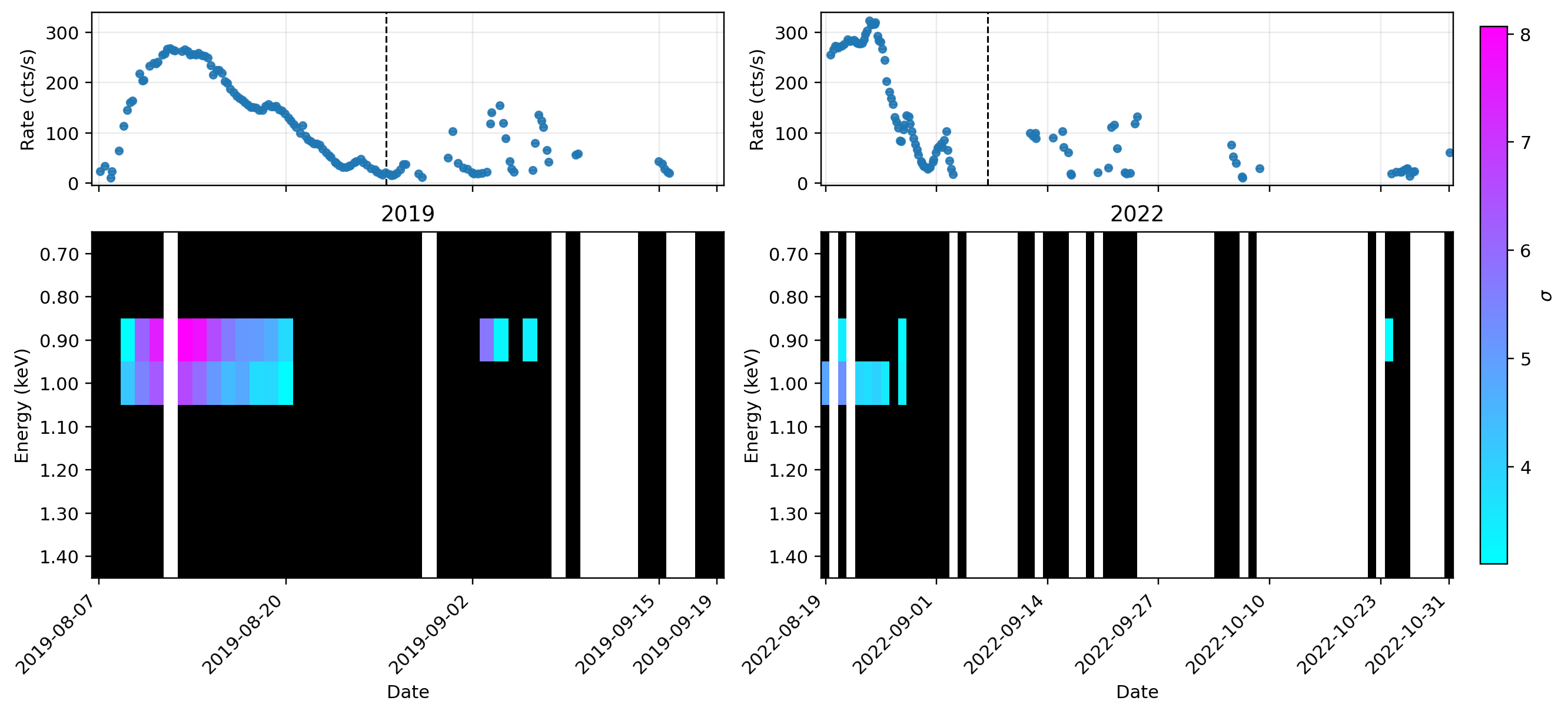}
    \caption{Search for a narrow $\sim$1~keV feature using fixed-energy Gaussian trials. 
For each \textit{NICER} observation, we add a zero-width Gaussian at a fixed energy to the baseline continuum and refit, freezing the line energy per trial (Section~\ref{sec:results}). 
A trial is counted as a detection if it improves the fit by $\Delta\chi^{2}\!\ge\!9$ (corresponding to 3$\sigma$ single-trial significance) compared to the no-line baseline and yields a line normalization $>\!5\times10^{-5}$~ph~cm$^{-2}$~s$^{-1}$. 
Symbols show detections as a function of time for the 2019 and 2022 campaigns; non-detections are omitted for clarity. 
Detections cluster at 0.90~keV and 1.0~keV, occurring near the outburst peaks and during brief reflaring intervals.}
    \label{fig:1kev}
  \end{figure*}

The temporal evolution of the feature in Figure \ref{fig:1kev} is most clearly reflected in the count rate panel of Figure \ref{fig:compplot}. To further investigate this apparent relationship, we computed the flux of each spectral component by convolving \texttt{CFLUX} with the components describing the primary continuum, the reflection spectrum, and the blackbody component over the modeled energy range. We then tested for a linear correlation between all combinations of model parameters,  including the Gaussian line strength, photon index, blackbody temperature, reflection fraction, and calculated component fluxes, using 20,000-iteration Monte Carlo simulations.

In each iteration, the parameters were resampled assuming Gaussian distributions centered on their measured values, with 1$\sigma$ uncertainties for quantities shown in Figure \ref{fig:compplot} and 3$\sigma$ uncertainties for the line strength and fluxes. For each iteration, we computed both the Pearson correlation coefficient and the errors-in-variables (Deming) regression slope. 
From the resulting distributions, we report the median correlation slopes, along with the central 99\% (3$\sigma$) intervals for cases where a significant correlation is detected, defined as $|r|\ge 0.7$ with confidence intervals that do not cross zero. 

During the 2019 outburst, the Gaussian line normalization exhibits a strong positive correlation with source brightness, as traced by the count rate (r=$0.8^{+0.1}_{-0.3}$). Consistent with this, similarly strong correlations are observed with both the primary continuum flux (r=$0.9\pm0.1$) and the blackbody flux (r=$0.8\pm0.1$), while a somewhat weaker but still significant correlation is present with the reflection flux (r=$0.7^{+0.1}_{-0.2}$). 

In the 2022 outburst, the correlation between the line strength and count rate remains comparably strong (r=$0.8^{+0.1}_{-0.3}$), as do the correlations with the primary continuum  (r=$0.9\pm0.1$) and the blackbody flux (r=$0.9\pm0.1$). In contrast, no statistically significant correlation is detected with the reflection flux, but large uncertainty on the correlation coefficient show that the data is not constraining (r=$0.1^{+0.5}_{-0.6}$). 

\begin{figure*}[bt]
    \centering
    \includegraphics[width=\linewidth]{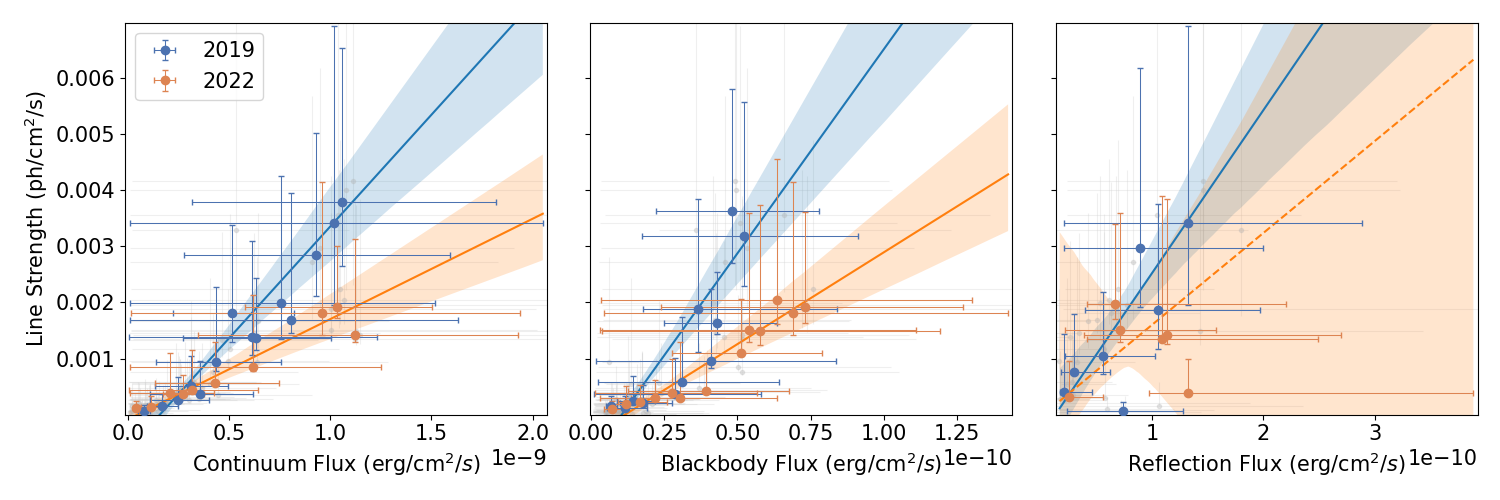}
    \caption{Strength of the $\sim$1~keV feature versus component flux. 
For all detections from the fixed-energy trials (Figure~\ref{fig:1kev}), we plot the Gaussian line normalization against the calculated flux for each of the \textit{NICER} observations. Colors distinguish outbursts (2019 in blue, 2022 in orange). Error bars are 3$\sigma$ on the normalization and fluxes.}
    \label{fig:linenorm}
\end{figure*}

To further examine these relationships, each outburst was binned into 15 equal-population intervals, and the Gaussian line normalization was plotted against the corresponding component fluxes (Figure \ref{fig:linenorm}). The binned Gaussian line normalizations track the variations in the primary continuum and blackbody fluxes across both outbursts, highlighting the systematic nature of these correlations. These trends reinforce the strong correspondence between the line strength and both the continuum and blackbody fluxes in each epoch. The Deming regression slopes show a much shallower relationship in 2022 compared to 2019, decreasing from $(4.0\pm0.3)\times10^{6}$ to $(1.8\pm0.2)\times10^{6}$ for the primary continuum and from $(7.3\pm0.5)\times10^{7}$ to $(3.3\pm0.3) \times10^{7}$ for the blackbody component. While a positive slope is measured for the reflection flux in 2019, the large uncertainties and lack of a significant correlation in 2022 indicate that this relationship is not robust.

These results demonstrate that the 1\,keV Gaussian line exhibits systematic correlations with both the primary continuum and blackbody fluxes, with variations in correlation strength between the 2019 and 2022 outbursts. The analysis presented here provides a quantitative framework for interpreting these differences, which we explore in detail in the following section.

\section{Discussion and Conclusions}\label{sec:disc}

We investigated the \textit{NICER} spectra of J1808 during its 2019 and 2022 outbursts. We found that the X-ray spectra are broadly well described throughout each outburst by a model with photoelectric absorption, a thermally Comptonized primary continuum, disk reflection, and a blackbody component, consistent with previous studies \citep{disalvo, Papitto2009, ludlam2018, hall25}. 

After establishing this baseline model, we introduced Gaussian lines to probe residuals near 1\,keV and characterized the feature phenomenologically. All lines that are significantly detected lie in the 0.9--1.0\,keV range, with stronger detections near 0.9\,keV in 2019 and near 1.0\,keV in 2022. We find that both the presence and strength of this feature correlate strongly with the continuum flux, blackbody flux, and reflection flux in 2019. In 2022, the correlations with continuum flux and blackbody flux are still significant, with Pearson coefficients comparable to those measured in 2019, although corresponding regression slopes are substantially shallower. The correlation with the reflection flux is highly unconstrained, and is therefore consistent with both weak or no correlation. The large uncertainties reflected the limited constraints on the 2022 dataset, and no trend with the reflection flux can be established. 

Before discussing possible astrophysical origins, it is important to discuss the possibility of instrumental effects. \citet{hall25} noted that the 1 keV feature is difficult to distinguish confidently from low-energy calibration uncertainties in current \textit{NICER} data. They also suggested that improvements in \textit{NICER}'s calibration since the early mission may contribute to differences between earlier and more recent studies. In the present work, however, the feature recurs during both the 2019 and 2022 outbursts, remains localized near 0.9-1.0 keV, and occurs preferentially near outburst peaks and during reflaring periods. The concentration of detections during certain periods suggests that the feature only becomes detectable during certain accretion conditions. Consistent with the observed correlations with continuum and blackbody flux, the absence of significant detections during the rise and decay phases may simply indicate that the feature weakens below our detection threshold of 3\,$\sigma$ at lower luminosities. We also looked for and did not find any indication of increased background during periods when the feature is not detected, nor evidence for optical loading or associated energy-scale calibration effects. While we cannot completely exclude a contribution from residual calibration uncertainties, we consider an instrumental origin very unlikely.

Several physical origins have previously been proposed for similar soft X-ray features. One proposed origin for similar features is burst-driven winds. Studies suggest that similar features in 4U~1820--30 arise from photospheric radius expansion during thermonuclear bursts \citep{strohmayer2019,mandal,yu25}. In J1808, this feature has also been detected during bursts \citep{casten23,bult}, indicating that burst-related processes could contribute under certain conditions. However, this scenario cannot account for the observed behavior of the feature throughout the outburst. This feature varies systematically with the persistent emission, rather than with the occurrence of thermonuclear bursts, indicating that it is tied to the accretion flow rather than transient burst events. 

While burst-driven winds are unlikely to explain the observed behavior of the feature, a related possibility is a disk wind. Ionized winds in neutron star low-mass X-ray binaries can produce soft X-ray features from light elements and Fe L-shell transitions \citep[e.g.,][]{munos2026,diaztrigo2006,ueda2004,ponti2014}, and could in principle generate emission or absorption near $\sim$1\,keV that scales with accretion rate, consistent with the observed correlations with continuum and blackbody flux. Indeed, thermal and radiation-assisted winds are expected to strengthen with increasing soft X-ray luminosity and disk temperature \citep{begelman1983,done2007,ponti2014}. However, even accounting for \textit{NICER}'s spectral resolution, winds are typically observed as structured, narrow, blueshifted absorption lines affecting multiple adjacent energy bins \citep{ueda2004,ponti2014,diaztrigo2016}, sometimes accompanied by P~Cygni--like profiles, rather than the smooth, unresolved excess seen here, which is confined to at most one to two adjacent \textit{NICER} energy bins and shows no coherent multi-bin structure, velocity signature, or systematic energy shifts. Moreover, disk winds in neutron-star systems are rarely observed as isolated emission features \citep{ponti2014,diaztrigo2016}. In addition, partial covering by neutral or weakly ionized material is disfavored, as it would introduce broad spectral curvature and absorption edges (e.g.\ O or Fe-L) rather than a localized excess. A wind contribution is therefore unlikely to be dominant, although it cannot be fully excluded. Instead, the smooth, flux-correlated nature of the feature and its close coupling to the reflection component in 2019 are more naturally explained by disk reprocessing.

In this framework, the correlation with reflection flux in 2019 naturally suggests that the feature is at least partially coupled to the reflection component of the spectrum. However, the weaker reflection coupling in 2022 implies that this connection is not stable across all accretion states, pointing to additional physical dependencies. Similar $\sim$1\,keV features have also been observed in other neutron star LMXBs, including Ser~X-1, GX~349+2, and 4U~1636--536, where higher-resolution spectroscopy suggests that they arise from blends of Fe\,L-shell and light-element transitions (e.g., O and Ne), rather than a single emission line \citep{cackett2010,ng2010,ludlam2018,schulz2001}. This supports an interpretation in which the feature reflects complex atomic structure in the inner accretion flow. Recent analyses of J1808 in 2019 and 2022 have also found that including reflection improves the modeling of this feature, consistent with (though not uniquely requiring) a disk-related origin \citep{dorsman26}. 

In J1808, the accretion flow is additionally shaped by the neutron star magnetic field, which truncates the inner accretion disk and channels material onto the magnetic poles \citep{Ghosh1979, Psaltis1999}. The observed $\sim$1\,keV feature may also arise from the dense ionized plasma near the magnetospheric truncation region, where hard X-rays from the accretion shock and hotspot irradiate the inner disk edge and magnetically confined accretion flow \citep{patruno2021}. Such a geometry naturally produces a dense reprocessing region with strong ionization gradients and enhanced soft X-ray line emission. The observed correlation between the feature strength and the blackbody flux is particularly suggestive in this context, as the blackbody component in AMXPs is generally associated with hotspot emission from magnetically channeled accretion \citep{Poutanen2003, Papitto2005, Papitto2013}. This correlation further suggests that the line-forming region is closely linked to the irradiated inner accretion flow rather than to a distant or static reprocessing medium.

High-density reflection models further support this picture. It has been shown that increasing the reprocessor density ($\log n_e \gtrsim 18$--$19\,\mathrm{cm}^{-3}$) can significantly alter the soft X-ray reflection spectrum relative to the standard assumption ($n_e=10^{15}$\,cm$^{-3}$), boosting thermal emission below $\sim$2\,keV \citep{Garcia2016}. At densities inferred for the accretion disk of J1808 during the 2019 outburst (log $n_e\approx19.1$ \citep{bruce}), these effects would enhance emission in Fe\,L and light-element complexes near $\sim$1\,keV \citep{cackett2010,ludlam2018}, providing a natural mechanism for a strong link between the observed line strength, continuum fluxes and the reflection flux. Within magnetically truncated accretion flows, such high densities may correspond to a more compressed inner disk and a denser transition region near the magnetospheric boundary during phases of enhanced accretion \citep{patruno2012}. The stronger and broader $\sim$1\,keV feature observed in 2019 is therefore consistent with more efficient irradiation and reprocessing in a denser inner accretion environment.

The observed differences between the 2019 and 2022 outbursts further support this scenario. While one would naively predict a close coupling of the observed 1\,keV line strengths with reflection flux for a reflection origin, our results show that this coupling weakens significantly in 2022, suggesting that the line-forming region responds differently as the accretion state evolves \citep{Garcia2016}. This is qualitatively consistent with the lower inferred disk density in 2022 (log\,$n_e\approx17.3$ \citep{bruce}). Such a density would suppress the X-ray line production, reducing the contribution of line emission to the soft reflected spectrum and thereby weakening the observed coupling between the feature and the reflection component. Within the AMXP framework, this may additionally indicate changes in the magnetospheric truncation radius and inner accretion geometry between outbursts, which would modify both the density structure and irradiation pattern of the inner flow. We note, however, that the density estimates discussed here are derived within the same spectral framework used in this work and are therefore likely model dependent.

The different evolution of the continuum photon index provides further evidence for changes in the inner accretion environment between the two outbursts. While the spectrum hardened as the source brightened in 2019, the photon index remained relatively stable around the 2022 outburst peak before softening during the decline. Both outbursts nevertheless remained in the hard state throughout \citep{baglio2020}. We note that the 2022 \textit{NICER} campaign began near the outburst maximum and therefore did not sample the rise, limiting a direct comparison of the spectral evolution during this phase. Although the observed trends suggest that the continuum evolution and the properties of the soft X-ray feature are governed by the same underlying changes in the inner accretion environment, the present \textit{NICER}-only analysis cannot determine whether this reflects changes in coronal properties, the disk structure, or both.

In summary, the 1\,keV feature in J1808 is recurrent and flux-dependent. Its correlation with the continuum and blackbody flux, together with differences between outbursts, show that it is sensitive to the conditions of the inner accretion flow and is consistent with a reflection origin, although residual instrumental contributions cannot be completely excluded. Our results further suggest that the feature may serve as a tracer of the interaction between the accretion disk and the neutron star magnetosphere in AMXPs. In the broader context of neutron star low-mass X-ray binaries, this interpretation is consistent with the mixed observational picture reported in the literature, where similar features are sometimes seen and sometimes not. The behavior observed in J1808 suggests that these different conclusions are likely due to variations in disk density, ionization, magnetic truncation geometry, and accretion state between sources and epochs, which have made a clear interpretation challenging.

Our scenario makes several testable predictions: if the feature is primarily driven by high-density reflection associated with the magnetically modified inner accretion flow, its strength should track changes in the soft continuum, hotspot emission, and inferred disk density, and it should vary systematically with accretion state both within and across outbursts. Variations in the magnetospheric radius may further alter the geometry and irradiation of the line-forming region, potentially producing the observed epoch-dependent behavior \citep{patruno2012}. Future broadband observations, pulse-phase-resolved spectroscopy, and self-consistent density-sensitive reflection modeling across multiple systems, particularly with high-resolution missions such as \textit{XRISM} and \textit{NewAthena}, will be essential to clarify whether similar $\sim$1\,keV features share a common origin or reflect system-dependent accretion conditions. Establishing the origin of the 1\,keV feature will provide key insight into the structure of magnetically truncated accretion flows in neutron star systems and clarify the nature of similar features across low-mass X-ray binaries.

\section*{Acknowledgments}\label{sec:acknowledgment}
We would like to thank the anonymous referee for their useful feedback and comments. JC acknowledges support from the Leverhulme Trust grant RPG-2023-240. This research has made use of data and/or software provided by the High Energy Astrophysics Science Archive Research Center (HEASARC), which is a service of the Astrophysics Science Division at NASA/GSFC. 

\bibliography{main}{}
\bibliographystyle{aasjournalv7}

\appendix\label{appendix}
The values held fixed in the continuum modeling are shown in Table \ref{tab:params}. These were obtained from the best-fit broad-band spectral analysis of the 2019 and 2022 outbursts of SAX J1808 outlined in \cite{bruce}. 
\begin{table*}[h]
    \captionsetup{justification=centering}
    \caption{Best-fit parameters for the combined broad-band spectral analysis of the 2019 and 2022 outbursts of SAX J1808 using \textit{NICER} and \textit{NuSTAR}. The model (\texttt{TBfeo} $\times$ [\texttt{relxillCp} + \texttt{bbodyrad} + \texttt{gaussian}])consists of photoelectric absorption  (including interstellar iron and oxygen abundance), a blackbody, a thermally Comptonized primary continuum, the reflection component, and a soft Gaussian component near 1 keV. The iron abundance is in units of solar iron abundance. See \cite{bruce}.}
     \label{tab:params}
     
    \setlength{\tabcolsep}{10pt}
    \centering
    \begin{tabular}{llcc}
        \hline
        Component & Parameter & 2019 & 2022  \\
        \hline
         tbfeo & N$_{\text{H}}$ (*10$^{22}$ cm$^{-2}$) & 0.23 & 0.17\\
         tbfeo & O & 1.57 & 1.34 \\
         tbfeo & Fe & 1.36 & 3.03 \\
         relxillCp & a & \multicolumn{2}{c}{0.3}\\
         relxillCp & kT$_{\text{e}}$ (keV) & \multicolumn{2}{c}{74.2}\\
         relxillCp & kT$_{\text{seed}}$ (keV)  & \multicolumn{2}{c}{0.01 (fixed)} \\
         relxillCp & Index & 2.05 & 2.21 \\
         relxillCp & R$_{\text{in}} (GM/c^2) $ & 16.1 & 12.3 \\
         relxillCp & R$_{\text{out}} (GM/c^2) $ & \multicolumn{2}{c}{1000 (fixed)}\\
         relxillCp & Incl (deg)  &\multicolumn{2}{c}{72}\\
         relxillCp & LogN (cm$^{-3}$) & 19.1 & 17.3 \\
         relxillCp & Fe abund & \multicolumn{2}{c}{1.0}  \\
         relxillCp & log$\xi$ & 1.65 & 1.82 \\      
         \hline
    \end{tabular}
\end{table*}

\end{document}